# Development and Characterization of a Precisely Adjustable Fiber Polishing Arm.

Michael P. Smith*[a], Sabyasachi Chattopadhyay[b], Andrew Hauser [a], Joshua E Oppor [a], Matthew A Bershady [a,b,c], Marsha J. Wolf [a]

[a]Washburn Astronomical Laboratories, University of Wisconsin-Madison, 475 N. Charter St., Madison, WI, USA 53706; [b]South African Astronomical Observatory, 1 Observatory Rd, Observatory, Cape Town, 7925, South Africa; [c]Department of Astronomy, University of Cape Town, Private Bag X3,Rondebosch 7701, South Africa

## ABSTRACT

The development of bare fiber or air-gapped microlens-fiber coupled Integral Field Units (IFUs) for astronomical applications requires careful treatment of the fiber end-faces (terminations). Previous studies suggest that minimization of fiber end face irregularity leads to better optical performance in terms of the diminishing effect of focal ratio degradation. Polishing has typically been performed using commercial rotary polishers with multiple gradually decreasing grit sizes. These polishers generally lack the ability to carefully adjust angular position and polishing force. Control of these parameters vastly help in getting a repeatable and controllable polish over a variety of glass/epoxy/metal matrices that make up integral filed units and fiber slits.

A polishing arm is developed to polish the fiber terminations (IFU, mini- bundles and v-grooves) of the NIR Fiber System for the RSS spectrograph at SALT. The polishing arm angular adjustments ensure the correct position and orientation of each termination on the polishing surface during the polish.

Various studies have indicated that the fiber focal ratio also degrades if the fiber end face comes under excessive stress. The polishing arm is fitted with a load cell to enable control of the polishing force. We have explored the minimal applicable end stress by applying different loads while polishing.

The arm is modular to hold a variety of fiber termination styles. The polishing arm is also designed to access a fiber inspection microscope without removing the fiber termination from the arm. This enables inspection of the finish quality at various stages through polishing process.

## 1. INTRODUCTION

Fiber performance is maximized when they are well terminated and carefully mounted and routed to minimize stress. It has been shown that the end-polish can have considerable impact on the total throughput and focal-ratio degradation (FRD; Eigenbrot et al. 2012). Specifically, Eigenbrot et al. (2012) found that there were significant performance gains for polishing down to surface-roughness delivered by 1 μm grit, where scratch and dig is uniform across the fiber surface at the 1μm level. One mitigation strategy is to bond fibers to commercially polished optics. This bonding offers the opportunity to AR-coat the incident surface (although this can be done for bare fiber arrays as well, e.g., Drory et al. 2015). However, this bonding is not always possible or desirable given fiber geometries, mechanical constraints, or optical design (e.g., in a telecentric pupil reimaging system). Therefore, it is desirable to have a routine method to polish fibers as singlets or in large arrays to have a high quality, uniform finish. Unfortunately, most commercial polishing stations do not provide ample control on polishing force to enable such a routine procedure to be developed.

To mitigate this shortfall and to overcome challenges we faced in polishing earlier generations of bare fiber IFUs, e.g., SparsePak (Bershady et al. 2004), HexPak (Wood et al. 2012) and ∇Pak (Eigenbrot & Bershady 2018), we have developed a custom polishing arm for use with an off-the-shelf Ultra Tec Polisher to enable precise polishing of fiber end terminations to be used in astronomical applications. The arm was developed to facilitate a more deterministic approach to polish v-groove blocks and the IFU for the SALT NIR Spectrograph (Sheinis et al. 2006; Wolf et al. 2014,

*mps@astro.wisc.edu

2018; Smith et al. 2018). The intension is that the features of the new arm would enable polishing procedures to be established and retained over time.

## 1.1 Design Elements

The primary requirements we introduced in our design included first and foremost (1) load-control in (2) a rigid structure that allowed the polishing specimen to be held at a well-defined (e.g., orthogonal) and reproduceable angle with respect to the polishing surface. Mechanical alignment in this context is essential. (3) We also required there be capability for in-situ inspection (via microscope) of the sample during polishing stages *without* removing the specimen from the mechanical structure. The latter stipulation is to preserve *exact* angular alignment and prevent differential faceting between inspections and grits. (4) We also desired to have the capability of moving the specimen to different (radial) locations on the polishing disk (platen) as well as different orientations with respect to the direction of polishing platen rotation. The range of location is motivated by the wish to use as much of the polishing disk area as possible; the range of angles is to avoid excessive groove structure during polishing in any one direction.

A standard feature of commercial polishers using rotating platens often includes the ability to facet the polishing surface and oscillate the specimen across the platen surface. While polishing non-orthogonal surfaces into fibers can be very useful, e.g., in application to curved spectrograph slits, the standard features for introducing facets do not offer the ability for polishing curved surfaces. Consequently, we did not consider implementing angular adjustment of the fiber polishing surface although this can be done via suitable fixturing. We also did not require motorization for specimen oscillation. This concession was largely driven by cost and complexity, although the ability to manually reorient the specimen is a requirement (see 4 above). As we will see, specimen oscillation turns out to be harmful in our experiments to achieving excellent end-finish at the finest grits.

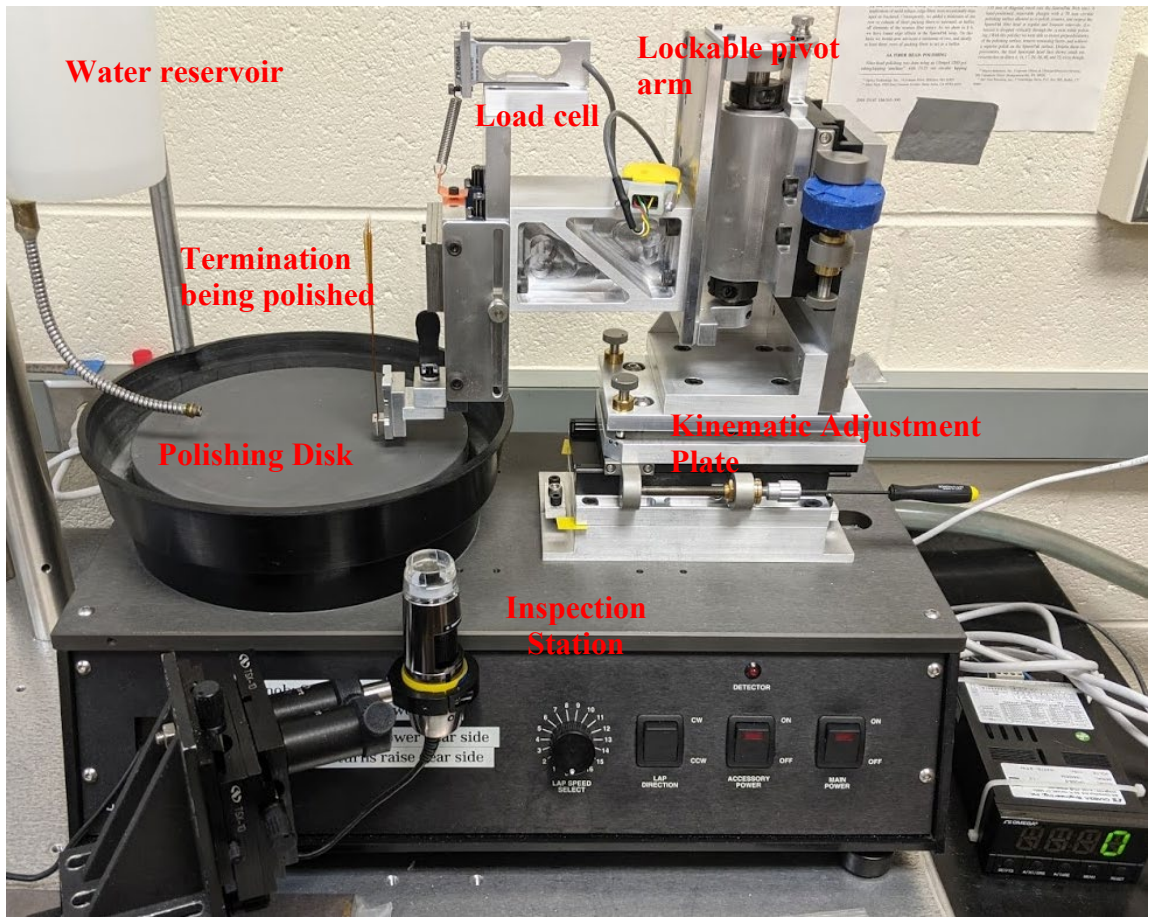

Figure 1. Photograph of the polishing arm setup showing the arm suspending a test v-groove termination above the polishing disk. The inspection microscope is in the foreground.

# 2. POLISHING ARM DESIGN

## 2.1 Overview

The upgraded polishing arm is intended to improve fiber polishing quality and efficiency by facilitating better control over key parameters. It is also intended to enable a more procedural approach to polishing that allows various lab staff and students to become proficient at polishing and allow procedures to be extended to other types of end terminations. The key design enhancements are a more precise angular alignment of the polishing plane, better polishing load control, and the ability to inspect the polished surface without removing it from the polisher.

## 2.2 Angular Alignment

The arm mechanics consist of two commercial orthogonal linear stages which enable x and z adjustment. The x adjustment sets the starting radial position of the polishing interface. The z-stage is adjustable for the nominal height of the part being polished. It also provides the fine adjustment of the polishing load force in combination with the load cell. A lockable pivot allows the arm to rotate and set the orientation of the termination specimen (fiber or fiber array) with respect to the direction of rotation of the polishing disk. The pivot also allows the termination to be rotated off the polishing disk to an inspection station without disturbing the angular alignment. The mechanical assembly is designed with solid mechanical sections and stages to be stiff and stable to minimize flexure during polishing and to accommodate larger end terminations and cables if needed.

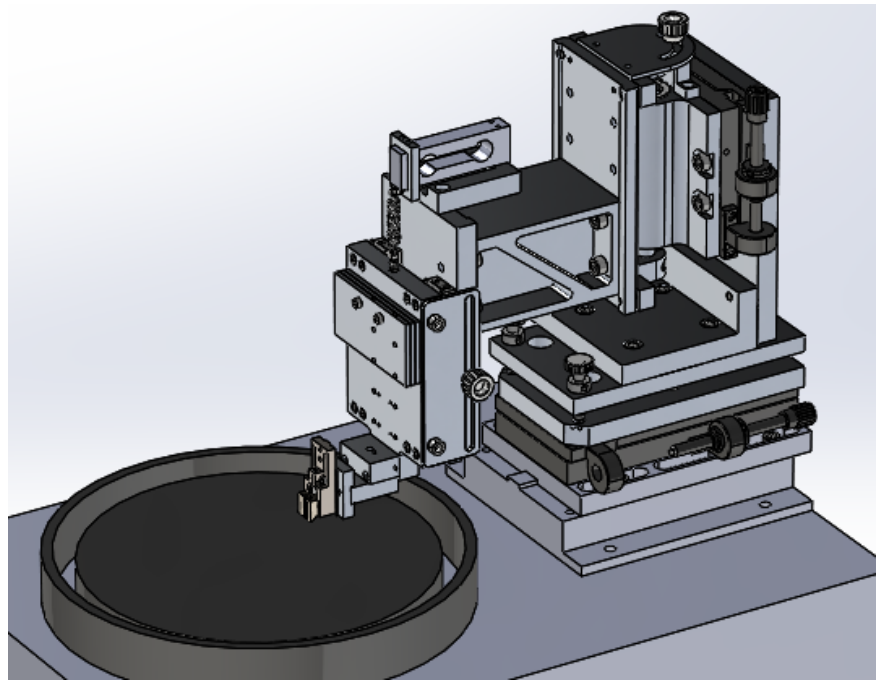

The kinematic tip tilt mount allows the arm's pivot axis and x-axis to be adjusted to be orthogonal to the polishing surface. This ensures a perpendicular polishing surface across the polishing film when the termination is keyed into the modular termination mount features on the mount interface. The angular adjustment is performed using a dial indicator to measure and zero any runout across the polishing surface. The mechanical structure is quite robust so that this alignment only needs to be done once, or after any disassembly of the arm mechanics or polishing platen.

Figure 2. Mechanical Model of the polishing arm showing the adjustable stages that enable adjustment of the polishing plane as well as precise positioning of the termination being polished

## 2.3 Load Control

Load control is achieved by a third linear stage that is mounted vertically at the end of the arm and allowed to float in tension with a spring attached to a load cell. The spring can be exchanged to alter the stiffnesses and the trim masses that are attached to the stage can be increase to control of the inertia response of the suspended mass. A small damper may be a good upgrade path should we experience any chatter, but to date this has not been evident.

The surface to be polished is then suspended above the polishing disk. Prior to contact being established, the load cell readout is zeroed out. The z-stage adjuster is then used to lower the polishing surface into contact with the polishing disk and the force (or pressure) of contact is established. The load cell and readout were inexpensive off-the-shelf units from Omega Engineering.

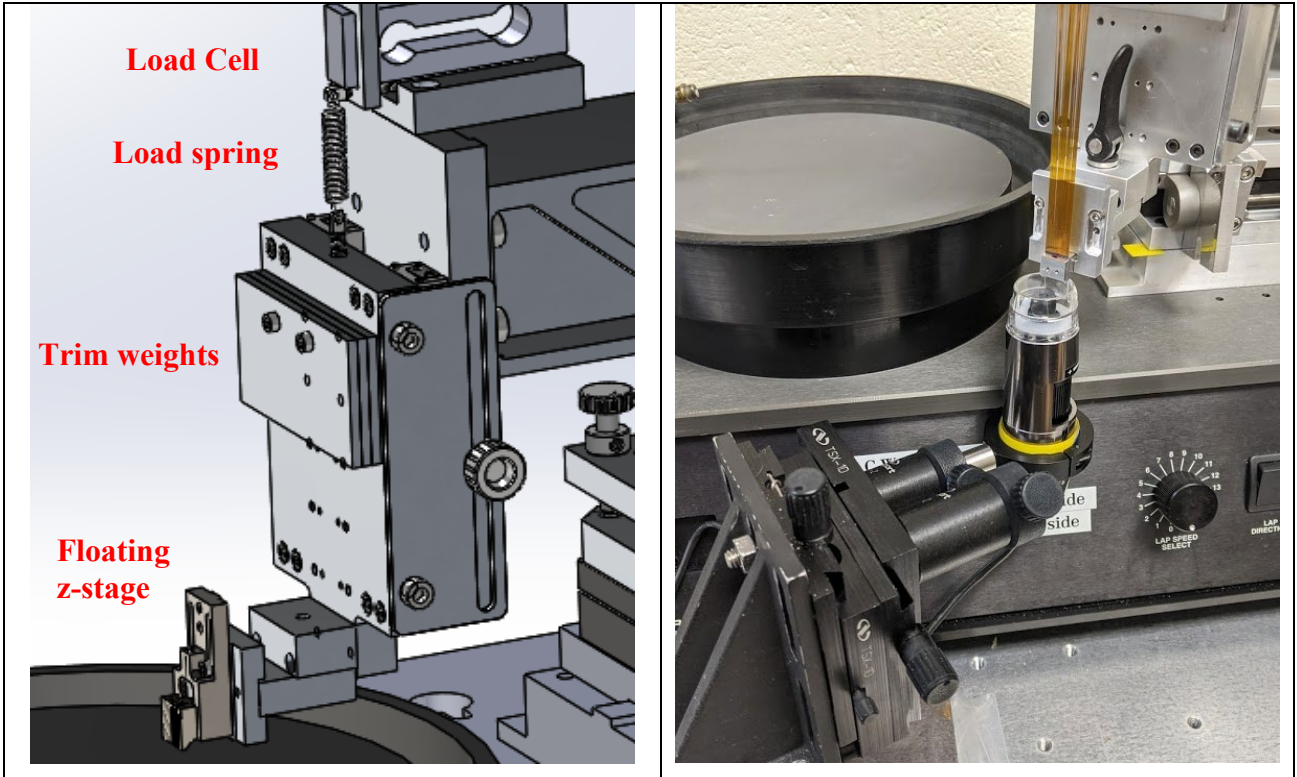

Figure 3. (Left) Model highlighting the load cell system on the new test arm. (Right) Photograph of the inspection station. The v-groove being polished here is rotated off the polishing disk and positioned above the Dino-Lite inspection microscope.

### 2.4 In-situ Inspection

By raising the termination and unlocking the swing arm the termination can be rotated off the polishing disk and positioned over the inspection station. The inspection station is a small upwards facing USB inspection microscope. We used a 450X microscope from Dino-Lite. The microscope is attached to a z-stage to allow for focus adjustment. We found this configuration to deliver necessary image quality to examine for dig and scratch features at the 1µm level. The short vertical microscope in this configuration is much easier to implement than trying to fold the light path with the addition of reimaging optics. A useful feature of the Dino-lite microscope is the ability to choose between circumferential and axial illumination. The different elimination methods show up different structures in the fiber which provide deeper insights into the quality of the polish as seen below.

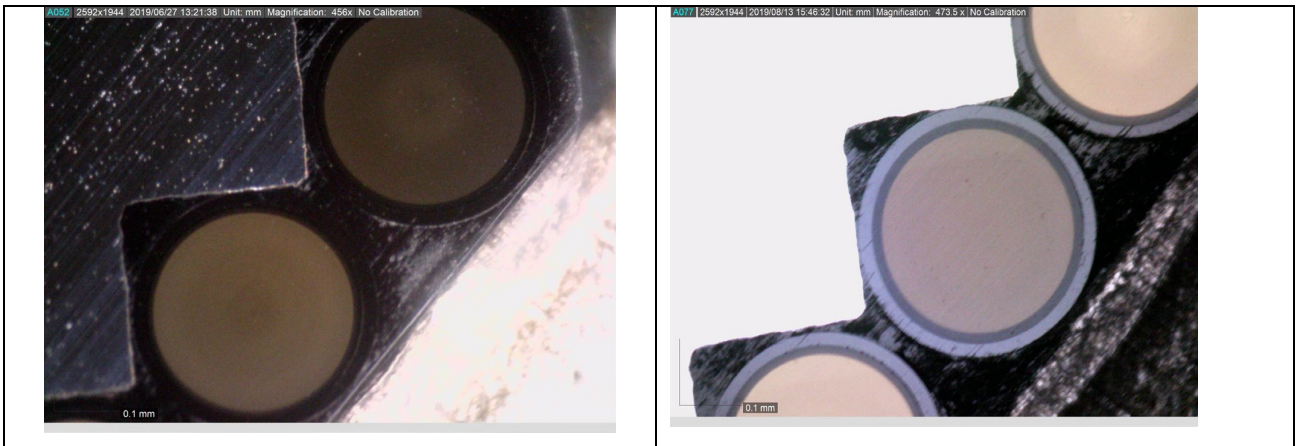

Figure 4. Comparison of circumferential illumination vs axial illumination of the microscope image of the polished fibers in a v-groove block.

# 3. POLISHING PROCEDURE AND RESULTS

## 3.1 Procedure

The starting point for our polishing procedure was based on our prior experience using the Ultra Tec polisher with the standard arm and aimed to take advantage of the addition functionality of the new arm describes above.

First the arm was leveled using the three adjustment screws on the kinematic base. Three extreme positions on the polishing disk were mapped out and probed using a dial indicator over the range of travel offered by the swing arm and x-stage. The kinematic mount was set in tip tilt to ensure the swing-arm/x-stage plane was parallel to the surface of the polishing disk. We later found that we were able to use the load cell as a sensitive touch indicator instead of the dial indicator.

To polish, we have settled on the following recipe: We start with 15μm grit polishing disks (30 μm if there is excessive material to remove) at around 5000 rpm, and decrease to 1 um grit at 1500 rpm, given in Table 1 below. Improvement in the surface roughness are evident going to 0.3 μm grit, but as earlier experiments have noted there is little gain in fiber performance (e.g., focal ratio degradation) when polishing beyond 1 μm grit (Eigenbrot 2012). As noted earlier, once the terminated fiber is mounted to the modular interface and is suspended above the polishing disk by the load cell spring, the load-cell readout is zeroed, and then the termination is slowly lowered to the polishing disk surface. At coarser grit, mechanical contact is easily heard, but this becomes difficult below 5 um grit size. Once the load cell reads a few grams, contact is secure, and a measure of the contact force can be made. The outstanding parameters in the polishing process are the contact force and polishing time per grit size.

We have established polishing times and force by experimentation with test terminations. In general, single fiber ferrules required ~10g of polishing force, v-groove blocks in the 100g range and larger IFU's slightly higher, given their relative characteristic sizes. We found that in general the ideal force scales with contact area such that the pressure is roughly 5 g $mm^{-2}$, and that *this pressure should remain the same at all grit*.

The table below shows the parameters that were found to be the best for two different styles of end terminations. The first was a single fiver ferrule with a 200μm fiber glued into a simple brass ferrule. The second was a v-groove block comprising of 25, 300 μm core fibers. In all these polishing runs, the termination being polished was positioned approximately 75 mm from the center of the 100 mm radius polishing disk for this particular calibration; in principle this can be used to scale disk rotation rate (rpm) with polishing radius to keep velocity constant.

| Single Fiber Ferrule (200μm fiber) | | | | V-Groove Block (300μm fiber) | | | |
| --- | --- | --- | --- | --- | --- | --- | --- |
| Grit (μm) | Speed (rpm) | Load (g) | Time (min) | Grit (μm) | Speed (rpm) | Load (g) | Time (min) |
| 15 | 5000 | 9 | 8 | 15 | 5000 | 90 | 20 |
| 5 | 4000 | 9 | 15 | 5 | 4000 | 90 | 35 |
| 3 | 2000 | 9 | 30 | 3 | 2000 | 90 | 45 |
| 1 | 1500 | 9 | 45 | 1 | 2000 | 90 | 55 |

Table 1. Polishing parameters for two different styles of end terminations that were found to give the best results.

Additionally, we were always careful to handle the polishing disks with gloves, ensure the disks were clean, and we used only distilled water. The polishing disks we used do not have adhesive backing and are held in place well by surface-tension for forces applied during polishing. It is essential to avoid air-pockets in the water adhesion process, particularly at finer grit.

## 3.2 Results

Figure 4 presents images taken of portions of v-groove blocks at various stages of polishing of 25 fibers in a stainless-steel V-groove block used to build a fiber pseudo-slit. The blocks were polished using different loads (listed at the top of each column). This visual inspection informed our conclusions summarized in Table 1.

One thing that we noted is that while it is tempting to move the specimen on the polishing film during polishing, such movements often lead to significant defects at any grit. Based on the appearance of these defects, movements under load leads to debris, likely from crumbling fractured regions scraping across the termination. In short, the lessons learned is not to move the specimen on the polishing disk while it is under load.

The most significant result from our development is that the quality of the polish is repeatable following our specific recipe.

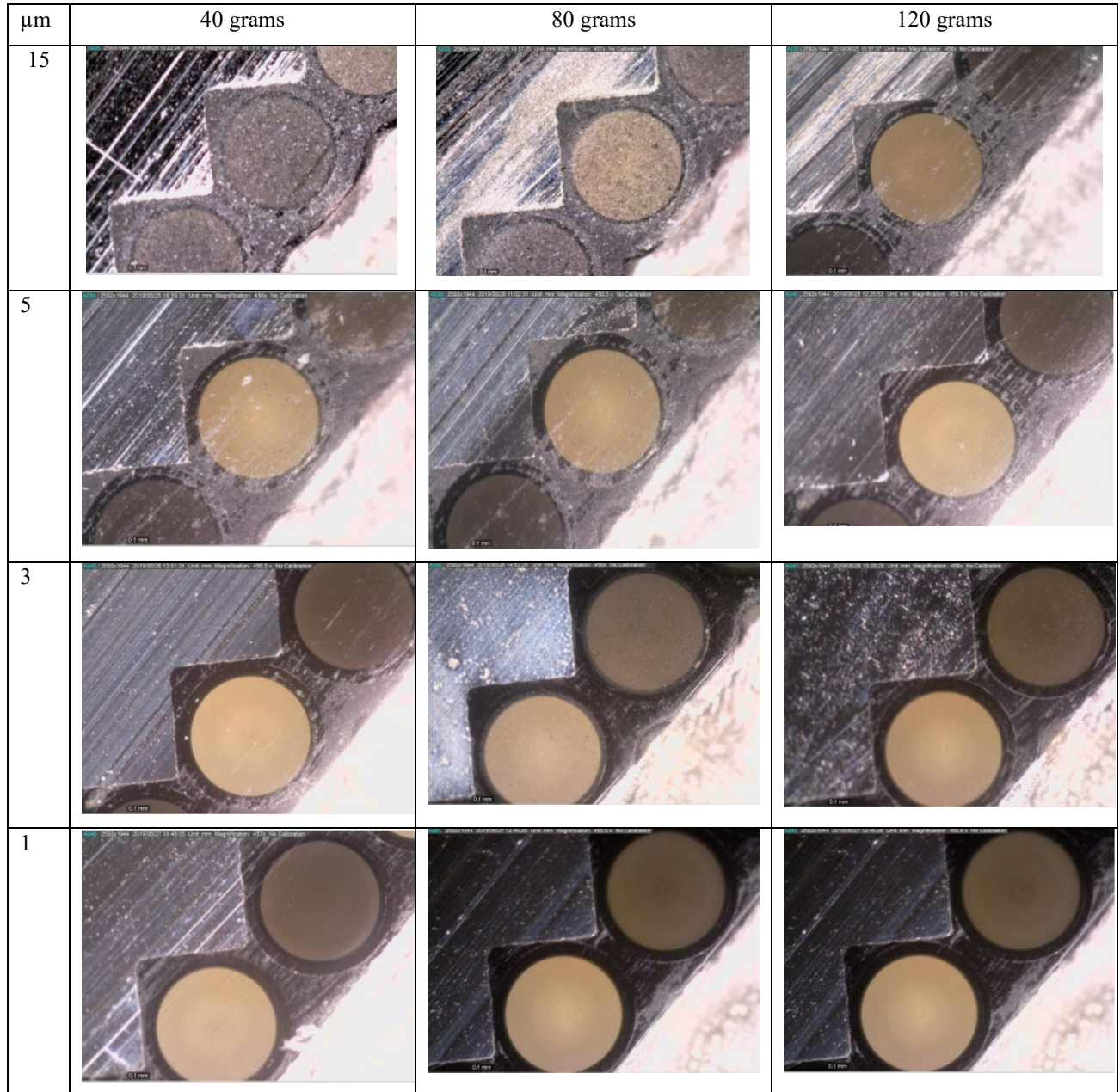

Figure 5. Photograph of a v-groove at various stages of polishing. Each column show a different loads setting as we step down the rows through decreasing polishing film grit.

## 4. CONCLUSION

The new polishing arm provides a simple and robust addition to the Ultra-Tec polisher used in our fiber labs at the University of Wisconsin-Madison as well as the South African Astronomical Observatory. The arm provides for a precise and stiff alignment of the termination to the polishing surface. The modular interface allows for quick and repeatable attachment of various types of end terminations. The load cell gives immediate feedback of polishing force and allows polishing procedures to be established and easily repeated.

Using the inspection station to monitor the progress of the polish and determining when to move to finer grit is extremely helpful. The ability to record images at different stages of polish enables detailed procedures to be written and followed and leads to repeatable results. The new arm will benefit both experienced operators in fine tuning polishing procedures on new and complicated fiber terminations and new operators at following procedures and achieving quality standards.

## REFERENCES

[1] Bershady, Matthew A., Andersen, David R., Harker, Justin, Ramsey, Larry W., and Verheijen, Marc A. W., "SparsePak: A Formatted Fiber Field Unit for the WIYN Telescope Bench Spectrograph. I. Design, Construction, and Calibration," PASP 116(820), 565 (2004).
[2] Drory, N., et al., "The MaNGA Integral Field Unit Fiber Feed System for the Sloan 2.5 m Telescope," AJ 149, 77 (2015).
[3] Eigenbrot, Arthur D., Bershady, Matthew A., and Wood, Corey M., "The impact of surface-polish on the angular and wavelength dependence of fiber focal ratio degradation," Proc. SPIE 8446, 84465W (2012).
[4] Eigenbrot, Arthur D., Bershady, Matthew A., "Vertical Population Gradients in NGC 891: I. $\nabla$Pak Instrumentation and Spectral Data," ApJ, 853,114 (2018)
[5] Sheinis, Andrew I., Wolf, Marsha J., Bershady, Matthew A., Buckley, David A. H., Nordsieck, Kenneth H., and Williams, Ted B., "The NIR upgrade to the SALT Robert Stobie Spectrograph," Proc. SPIE 6269, 62694T (2006).
[6] Smith, Michael P., Wolf, Marsha J., Bershady, Matthew A., Adler, Douglas P., Jaehnig, Kurt P., Koch, Ron J., Mulligan, Mark P., Oppor, Joshua E., Percival, Jeffrey W., Aydinyan, Nelli, Hauser, Andrew S., Ruder, Elijah. "A Near Infrared Integral Field spectrograph (NIR) for the Southern African Large Telescope (SALT): mechanical design" Proc. SPIE. 10702, (2018)
[7] Wolf, Marsha J., Mulligan, Mark P., Smith, Michael P., Adler, Douglas P., Bartosz, Curtis M., Bershady, Matthew A., Buckley, David A. H., Burse, Mahesh P., Chordia, Pravin A., Clemens, J. Christopher, Epps, Harland W., Garot, Kristine, Indahl, Briana L., Jaehnig, Kurt P., Koch, Ron J., Mason, William P., Mosby, Gregory, Nordsieck, Kenneth H., Percival, Jeffrey W., Punnadi, Sujit, Ramaprakash, Anamparambu N., Schier, J. Alan, Sheinis, Andrew I., Smee, Stephen A., Thielman, Donald J., Werner, Mark W., Williams, Theodore B., and Wong, Jeffrey P., "Project status of the Robert Stobie spectrograph near infrared instrument (RSS-NIR) for SALT," Proc. SPIE 9147, 91470B (2014).
[8] Wolf, Marsha J., Adler, Douglas P., Bershady, Matthew A., Jaehnig, Kurt P., Koch, Ron J., Mulligan, Mark P., Oppor, Joshua E., Percival, Jeffrey W., Smith, Michael P., Aydinyan, Nelli, Hauser, Andrew S., Ruder, Elijah., "A near infrared integral field spectrograph for the Southern African Large Telescope (SALT)," Proc. SPIE 10702-97 (2018)
[9] Wood, Corey M., Bershady, Matthew A., Eigenbrot, Arthur D., Buckley, Scott A., Gallagher, John S., Hooper, Eric J., Sheinis, Andrew I., Smith, Michael P., and Wolf, Marsha J., "HexPak and GradPak: variable-pitch dual-head IFUs for the WIYN 3.5m Telescope Bench Spectrograph," Proc. SPIE 8446, 84462W (2012).